\documentclass[pr,twocolumn,amsmath,nofootinbib,longbibliography,floatfix]{revtex4-1}
\usepackage{color}
\usepackage[dvipsnames]{xcolor}
\usepackage{graphicx}
\usepackage{bm}
\usepackage{float} 
\usepackage{mathrsfs}
\usepackage{amsthm}
\usepackage{dcolumn}
\usepackage{amssymb}
\usepackage{amsfonts}
\usepackage{hhline}
\usepackage{bbold}

\usepackage{ifpdf}
\ifpdf
   \usepackage[unicode=true,colorlinks=true]{hyperref}
\fi

\newcommand{\KP}{${\bm k} \cdot{ \bm p}$}

\newcommand{\ket}[1]{\left\vert #1 \right\rangle}

\renewcommand\vec[1]{\ensuremath\boldsymbol{#1}}

\begin{document}

\title{Valley and spin splittings in PbSe nanowires}

\author{I.D.~Avdeev$^1$}
\author{A.N.~Poddubny$^1$}
\author{S.V.~Goupalov$^{1,2}$}
\author{M.O.~Nestoklon$^1$}
\affiliation{$^1$~Ioffe Institute, St. Petersburg 194021, Russia\\
$^2$~Department of Physics, Jackson State University, Jackson MS 39217, USA}

\begin{abstract}
  We use an empirical tight-binding approach to calculate electron and hole states in [111]-grown PbSe nanowires.
  We show that the valley-orbit and spin-orbit splittings are very sensitive to the atomic arrangement 
  within the nanowire elementary cell and differ for [111]-nanowires with microscopic 
  $D_{3d}$, $C_{2h}$ and $D_{3}$ symmetries. 
  For the nanowire diameter below 4~nm the valley-orbit splittings become comparable with 
  the confinement energies and the $\bm k\cdot\bm p$ method is inapplicable. 
  Nanowires with the $D_{3}$ point symmetry having no inversion center exhibit giant spin 
  splitting $E = \alpha k_z$, linear in one-dimensional wave vector $k_z$, with 
  the constant $\alpha$ up to 1~eV$\cdot$\AA.

\end{abstract}


\maketitle

\section{Introduction}

Lead chalcogenide nanostructures are widely used for optoelectronics 
applications, including infrared detectors~\cite{Sukhovatkin09}, 
solar cells\cite{Sargent09,Semonin11},
and light emitting diodes~\cite{Garuge08,Sun13}.
Most of the works are devoted to zero-dimensional 
nanocrystals~\cite{Sukhovatkin09,Sargent09,Semonin11,Garuge08,Sun13,Efros2005}. 
However, quasi-one-dimensional lead chalcogenide nanowires (NWs) and nanorods are 
offering more flexibility.
In particular, enhanced multiple exciton generation 
efficiency~\cite{Melinger2011,Gesuele2012,Klimov2013,Davis15} and suppressed 
Auger recombination rates \cite{Htoon03,Padilha13,Siebbeles2013} have been experimentally demonstrated in PbSe NWs.  PbSe NWs can be grown using methods of colloidal chemistry~\cite{Lifshitz2003,Murray2005,Murray2010,Wise2011,Tischler2013,Tischler2015}. The approach   based on the oriented attachment of PbSe nanocrystals \cite{Murray2005} enables  controllable growth of NWs along the [100], [110] and [111] axes with straight, zigzag, helical, branched, and tapered shape. 
Growth of high-quality 
monocrystalline PbSe [100] nanorods with homogeneous size distributions using a catalyst-free, one-pot, solution chemistry method has been reported in Refs.~\cite{Murray2010,Wise2011}. This technique offers control over the nanorod aspect ratio in the range from 1 to 16~\cite{Tischler2013,Tischler2015}.
Energy spectrum of the nanowires can be probed optically~\cite{Bartnik10,Wise2011,Stroud2011,Tischler2014}. Simple stationary linear absorption spectroscopy easily probes the effect of the nanorod size \cite{Stroud2011}, and aspect ratio \cite{Tischler2013} on the fundamental energy  gap. More detailed information might be accessible using the nonlinear and transient  optical setups~\cite{Padilha2010,Tischler2014}.

While the basics of the electron structure of the nanowires is well understood, the valley structure of the states 
has not been fully investigated. 
The multi-valley band structure represents a challenge for the modeling, as one has to deal 
with electronic states, originating from different valleys, whose degeneracies are very sensitive to the 
microscopic properties and symmetry
of a nanostructure. The symmetry of a nanowire made of a crystalline material with the rock salt crystal lattice, having band extrema in four $L$ points of the Brillouin zone, is determined by several major factors. 

The first of these factors is the nanowire growth direction. In most theoretical papers~\cite{Abhijeet11,Bartnik10}, NWs grown along the two high-symmetry directions, [100] and [111], are considered. 
Theoretical works based on the effective mass method~\cite{Bartnik10,Goupalov11,Goupalov13} treat the valleys as independent. However, if the valley mass anisotropy is taken into account~\cite{Bartnik10}, this description becomes sensitive to the growth direction.
In the absence of intervalley coupling, the electronic states in [100] NWs remain 4-fold valley degenerate while in [111] NWs the longitudinal [111] valley state becomes split-off by the effective mass anisotropy.

The second factor is the shape of the NW cross-section. For a theoretical model, it is natural to choose it in such a way that it does not reduce the overall symmetry of the nanowire. The simplest choice is the shape which is as close to a circle as possible. 
The presence of the NW surface leads to mixing of the valley states (inter-valley coupling), which results in energy splittings~\cite{Ohkawa78,Nestoklon06}. 
The structure of electron energy levels resulting from the valley splittings may be determined from the overall nanowire symmetry which is the combination of the $O_h$ symmetry of the crystal lattice and the symmetry of the NW structure potential. 
The effective mass method may be extended to account for the valley degree of freedom, so that it can be used to predict the degeneracies of resulting energy eigenstates and their transformation properties. Such an extension of the effective mass method and symmetry analysis of the inter-valley coupling is one of the aims of the present paper. However, a full quantitative analysis of the inter-valley coupling requires computational methods capable of accounting for the atomistic structure of the nanowire. Below we employ the empirical tight-binding method to perform such an analysis.
This method has proved to be effective in description of the inter-valley coupling in lead chalcogenide nanocrystal quantum dots~\cite{Poddubny12} and in nanowires with a rectangular cross-section~\cite{Abhijeet11}.
Contrary to {\it ab-initio} techniques \cite{Wrasse2011,Wrasse13}, that are as yet
limited to the NWs thinner than 30~\AA~ \cite{Wrasse2011,Wrasse13}, the tight-binding method can easily handle NWs with large diameters. An empirical tight-binding description also naturally accounts for the relativistic nature of Pb atoms that induces strong spin-orbit interaction, representing a challenge for the {\it ab-initio} approaches~\cite{Hummer07,Svane10}.

\begin{figure*}[tbp]
  \centering{\includegraphics[width=\linewidth]{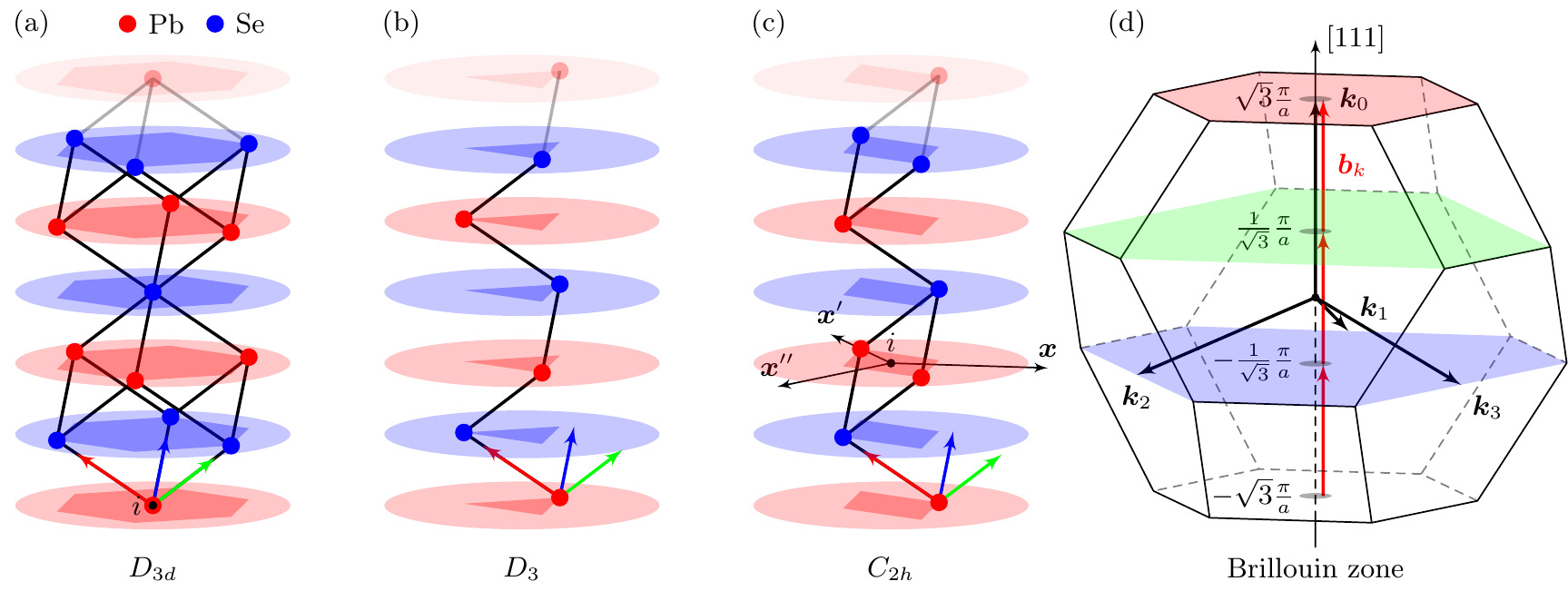}}
	\caption{(color online) Panels (a), (b), (c) show elementary cells of three smallest NWs with the point group symmetry $D_{3d}$, $D_3$, and $C_{2h}$, respectively. Cations (Pb) are shown by red dots, and anions (Se) by blue dots. Color arrows (RGB for XYZ) indicate the crystallographic coordinate system with the origin at the point $(0,0,0)$. Inversion center of the $D_{3d}$ NW is at the $(0,0,0)$ cation, while in the $C_{2h}$ NW the inversion center is chosen at the $a(\frac{1}{2},\frac{1}{4},\frac{1}{4})$ point in the crystallographic coordinate system.	
	Panel (d) shows the bulk PbSe Brillouin zone. Wave vectors $\vec{k}_i$ \eqref{eq:valley_def} are 
        shown as black arrows. The colored cross sections are formed by (111) planes which pass through
        the points $\vec{k}_1$, $-\vec{k}_1$, $\vec{k}_0$.
        Projections of these vectors to the [111] axis are indicated by tick marks on the [111] axis. 
		} 
		\label{fig:w_cells_BZ}
\end{figure*}

Even when the nanowire growth direction and the shape of its cross-section are chosen, the symmetry of the nanowire is not fully determined. For a nanowire with a cylindrical shape, the position of the axis of this cylinder within a unit cell of a bulk semiconductor is yet to be specified. It turns out that not only is the inter-valley coupling sensitive to the choice of this position, but entirely new physical consequences such as appearance of spin-dependent splittings of nanowire energy levels can also result. From this point of view, the [111] grown NWs appear to be more interesting, and we will restrict our consideration to these NWs. Analysis of the [100] grown NWs will be undertaken in a separate publication.

The paper is organized as follows. In Section~\ref{sec:wire_structure} 
we describe the system and explain the microscopic symmetry of the NWs. 
In Section~\ref{sec:kp} we describe the extension of the effective mass method that takes into account combinations of the valley states. We formally consider an ideal cylindrical nanowire, but restrict ourselves to the symmetry operations of the true nanowire symmetry group specified in Section~\ref{sec:wire_structure}. This allows us to 
predict the degeneracies of eigenenergies and transformation properties of eigenfunctions
resulting from the valley-orbit and spin-orbit splittings.
In Section~\ref{sec:TB} we explain the tight-binding approach and its application 
to lead chalcogenides NWs. We describe a technique for detailed analysis of 
the tight-binding wave functions in both real and reciprocal spaces. 
In Section~\ref{sec:results} we discuss the results of the tight-binding calculations 
and compare them with the effective mass consideration. 
In Section~\ref{sec:conclusion} the conclusions are drawn.

\section{Wire structure in real and reciprocal spaces}
\label{sec:wire_structure}

In this section we analyze the possible microscopic symmetries of the 
PbSe nanowires grown along the [111] direction.

Bulk PbSe has the rocksalt (fcc) crystal lattice with the $O_h$ point group. 
An elementary cell of a NW grown along [111] direction contains three Pb 
and three Se (111) atomic layers, 
as shown in Fig.~\ref{fig:w_cells_BZ}. 
NWs have one-dimensional periodicity along the growth direction characterized by the translational vector
\begin{equation} \label{eq:T_wires}
  \vec{T} = a(1,1,1) \:,
\end{equation} 
where $a$ is the PbSe lattice constant.
A nanowire is ``carved out'' of the
bulk crystal 
along a cylindrical surface with an axis parallel to
the [111] direction. 
Some possible arrangements of atoms in the elementary cells of resulting nanowires are shown in Fig.~\ref{fig:w_cells_BZ}.
They differ by the position of the NW axis with respect to the unit cell of a bulk crystal leading to different point symmetries of the resulting NWs.  In particular, Fig.~\ref{fig:w_cells_BZ} shows
three typical NWs with the center axis  
passing through the atom ($D_{3d}$, Fig.~\ref{fig:w_cells_BZ}a), 
between atoms ($D_{3}$, Fig.~\ref{fig:w_cells_BZ}b) or through the middle of a chemical bond ($C_{2h}$, Fig.~\ref{fig:w_cells_BZ}c). In the case of the $D_{3d}$ NWs we choose the inversion center at the $(0,0,0)$ cation. 
 For the $C_{2h}$ NWs, the inversion center is chosen at $a(\frac{1}{2}, \frac{1}{4}, \frac{1}{4})$. 
Note that the crystallographic coordinate system is used only to define NW directions, their inversion centers and reciprocal lattice vectors.
Below we use the coordinate system with the following axes
\begin{equation}
	\label{eq:wire_axes}
	\begin{array}{l c r }
		\vec{x} \parallel  [\bar{1}10],~ &
		~\vec{y} \parallel [\bar{1}\bar{1}2],~ &
		~\vec{z} \parallel [111]
	\end{array}
\end{equation}
and, for the $D_{3d}$ and $C_{2h}$ NWs, the origin of this coordinate system is set to the inversion center.

Now we discuss the relations between the one-dimensional reciprocal space of a nanowire and the three-dimensional reciprocal space of the parent bulk material.
The conduction band minima and valence band maxima in bulk lead chalcogenides are located in the four inequivalent $L$ 
valleys. The first Brillouin zone is schematically shown in Fig.~\ref{fig:w_cells_BZ}d.
The point group of the wave vector in each $L$ valley is $D_{3d}$ with its $C_3$ axis directed along the valley. 
Four valleys $L_{\nu}$ correspond to the wave vectors $\vec{k}_{\nu}$ ($\nu=0,1,2,3$), 
which we choose as
\renewcommand{\arraystretch}{1.5} 
\begin{equation}
	\label{eq:valley_def}
	\begin{array}{l r}
		\vec{k}_0 = \frac{\pi}{a}(~~1,~~1,~~1)\:, & \vec{k}_1 = \frac{\pi}{a}(-1,-1,~~1)\:, \\
		\vec{k}_2 = \frac{\pi}{a}(~~1,-1,-1)\:, & \vec{k}_3 = \frac{\pi}{a}(-1,~~1,-1)\:.
	\end{array}
\end{equation}

We also choose the basis vectors of the bulk reciprocal lattice as 
\begin{equation}
	\label{eq:rlvs}
	\vec{b}_1=-2\vec{k}_3\:,~~\vec{b}_2=-2\vec{k}_1\:,~~\vec{b}_3=-2\vec{k}_2 \:.
\end{equation}
In the one-dimensional reciprocal space of a nanowire, the quantum-confined states, originating from the electron and hole states at the $L$ points in the 3D bulk Brillouin zone, are 
located at the projections of the $\vec{k}_{\nu}$ onto the 
wire one-dimensional Brillouin zone  defined by the vector 
\begin{equation}
	\label{eq:1d_BZ_new}
	\vec{b}_k = \frac{2 \pi}{|T|} \: \frac{1}{\sqrt{3}}(~1,~1,~1) \:,
\end{equation}
whose length $b_k = 2\pi/a\sqrt{3}$ is three times shorter than the length of the basis reciprocal lattice vectors.

To illustrate the relation between the 1D and 3D Brillouin zones, in 
Fig.~\ref{fig:w_cells_BZ}d we show three 1D Brillouin zones embedded into the 3D 
bulk Brillouin zone by red arrows. The
three cross section planes which cut the 3D Brillouin zone at the projections of the valleys $\pm\vec{k}_{0,1,2,3}$
onto the $[111]$ axis are painted in Fig.~\ref{fig:w_cells_BZ}d in red, green, and blue. The wave vectors $\vec{k}_{0,1,2,3}$ are shown by black arrows. 
The distance between the cross sections is equal to the length of the 1D Brillouin zone, 
which means that the states originating from the four inequivalent $L$ valleys become equaivalent in the $[111]$ NWs.

It is convenient to map all the
inequivalent $L$ valleys into a single (111) plane. 
The natural choice is the plane passing through the $\vec{k}_0$ point.
In order to  map the valleys into this plane, we add 
the reciprocal lattice vectors $\vec{b}_1$ and $\vec{b}_2 + \vec{b}_3$
to the vectors from the ``green'' and ``blue'' planes in Fig.~\ref{fig:w_cells_BZ}, respectively. 
Below we will adhere to this scheme and, instead of addressing vectors from the three
planes confined within the first 3D Brillouin zone, we will use equivalent vectors 
from the single plane.
The planar translation vectors 
can be expressed via the reciprocal lattice vectors
\begin{equation}
	\label{eq:K_SC_T}
	\vec{T}_{1}^{2D} = \vec{b}_1 - \vec{b}_2, ~~ \vec{T}_{2}^{2D} = \vec{b}_3 - \vec{b}_2\:.
\end{equation}

\section{Effective mass approach}
\label{sec:kp}

The \KP\ effective mass model to describe the electron energy spectrum near the extrema of the conduction and valence bands in a given $L$-valley of a bulk lead chalcogenide compound (PbS, PbSe, PbTe, etc.) was proposed by Dimmock and Wright~\cite{Dimmock64}. 
The isotropic version of this method which neglects the mass anisotropy 
was used to calculate the states in spherical nanocrystals~\cite{Kang97},
cylindrical nanowires~\cite{Goupalov11}, and cylindrical nanorods~\cite{Goupalov13}. 
Bartnik {\it et al.}~\cite{Bartnik10} applied an axially symmetric model which may be 
obtained by ``averaging'' the mass anisotropy in the directions, perpendicular to the 
nanowire growth direction, to describe quantum confined 
states in NWs. 
In what follows we further extend this approach to account for the mixing of the valley states. 
This enables us to classify the electron states in a nanowire with a given point symmetry (cf. Fig.~\ref{fig:w_cells_BZ})
with respect to their transformation properties. 

\begin{figure}[tbp]
  \centering{\includegraphics[width=\linewidth]{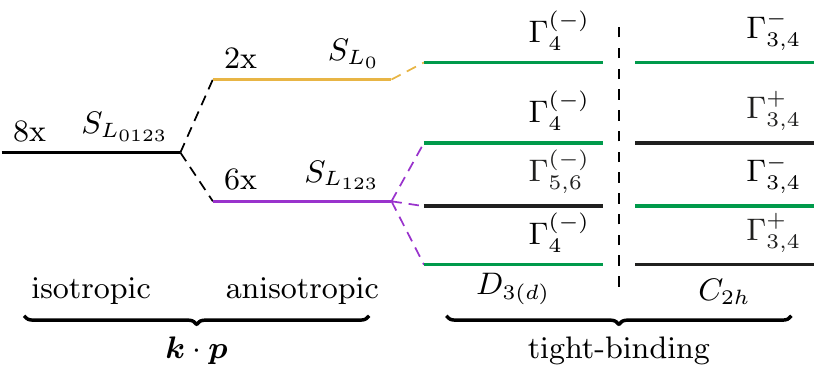}}
  \caption{Scheme of the lowest electron energy levels in $D_{3(d)}$ and $C_{2h}$ [111]-grown NWs.
  Left to right: isotropic \KP\ method; in \KP\ method with account of mass anisotropy; and with the valley splitting. 
  States in $D_3$ NWs have the same structure as in $D_{3d}$ without a certain parity. 
  } 
\label{fig:levels_scheme}
\end{figure}
 Figure~\ref{fig:levels_scheme} qualitatively  illustrates the degeneracies 
of the lowest electron levels in the conduction band of $D_{3(d)}$ and $C_{2h}$ symmetry
NWs grown along the [111] direction. At $k_z=0$ the level is 8-fold (4 valleys $\times$ 2 spin) degenerate within the isotropic effective mass model~\cite{Goupalov11}. The axially symmetric effective mass model~\cite{Bartnik10} partially lifts this degeneracy by splitting this level into the 2-fold degenerate level originating from the longitudinal $L_0$ valley and the 6-fold degenerate level formed by the states from the inclined valleys. Both the symmetry analysis in the framework of the \KP\ method presented below and the tight-binding calculations show that the states originating from the 3 inclined valleys are further split into three 2-fold degenerate levels. We label them in accordance with the irreducible representations of the point group of a NW using the Koster notation~\cite{Koster}.

\subsection{Bulk Hamiltonian}
\label{sec:bulk_KP}
The isotropic \KP~Hamiltonian written in the basis of the spinors and the Bloch functions in
the longitudinal valley $L_{0}$,
\begin{equation}
	\label{eq:0_basis}
	\vec{\mathcal{E}}_0 = \left(\ket{L_{0}^c}\ket{\uparrow},~ \ket{L_{0}^c}\ket{\downarrow},~ \ket{L_{0}^v}\ket{\uparrow},~ \ket{L_{0}^v}\ket{\downarrow}\right)\:,
\end{equation}
takes the form~\cite{Goupalov11,Dimmock64,Kang97} 
\begin{equation}
	\label{eq:H}
	H =
	\left[
		\begin{matrix}
			\left( \frac{E_{g}}{2} - \alpha_{c}\Delta \right) & -i P \left(  \vec{\sigma} \vec{\nabla} \right) \\
			i P \left( \vec{\sigma} \vec{\nabla} \right) & - \left( \frac{E_{g}}{2} - \alpha_{v}\Delta \right)
		\end{matrix}
	\right] \:,
\end{equation}
where $\alpha_c,~\alpha_v$ are combinations of remote band mass parameters, $P$ is the interband momentum matrix element. 

Usually the conduction band Bloch function $\ket{L_0^{c}}$ is odd and the valence band Bloch function$\ket{L_0^{v}}$ is even, however, we use the $\ket{L^{c,v}}$ notation instead of commonly used $\ket{L_0^{\mp}}$ (or simply $\ket{L^{\mp}}$) because the parities of the Bloch functions depend on the position of the inversion center, which is different in $D_{3d}$ and $C_{2h}$ NWs, see Fig.~\ref{fig:w_cells_BZ}.

It is worth to note that, for $k_z=0$, the anisotropic model of Ref.~\cite{Bartnik10} reduces to the isotropic one with renormalized parameters. Our \KP\ calculations will be limited to this case.
The \KP\ parameters can  be extracted from the band structure calculated using the tight-binding approach: 
$P^l=3.8859$ eV$\cdot$\AA,  $\alpha_c^l=22.9078$ eV$\cdot$\AA$^2$, $\alpha_v^l=27.5727$ eV$\cdot$\AA$^2$ for longitudinal valley $L_0$ and $P^i=3.5943$ eV$\cdot$\AA,  $\alpha_c^i=15.9308$ eV$\cdot$\AA$^2$, $\alpha_v^i=17.9751$ eV$\cdot$\AA$^2$ for inclined $L_{123}$ valleys, with the zero-temperature band gap $E_g=0.2129$ eV.

\subsection{Effective mass model for nanowire states originating from a single valley}

\subsubsection{Main equations}
First we construct the eigenstates of the Hamiltonian~\eqref{eq:H} at the longitudinal valley $L_0$.
The Hamiltonian Eq.~\eqref{eq:H} commutes with the total angular momentum operator~\cite{Kang97} 
\begin{equation}
	\label{eq:J}
	\hat{\vec{J}} = \hat{\vec{L}} + \hat{\vec{\Sigma}},
\end{equation}
where $\hat{\vec{\Sigma}}=\vec{\sigma}/2$. It is convenient to find solutions in the 
NW as eigenfunctions of the 
component, $\hat{J}_{z}$, of the total angular momentum along the growth direction. 
The eigenvalues of $\hat{J}_{z}$ are denoted by $m$.
Equation~(\ref{eq:H}) has  four linearly independent solutions  finite at $\rho = 0$, where $\rho$ is the in-plane radial coordinate of the cylindrical coordinate system. 
The boundary condition (which we choose in the form of vanishing wave functions at the surface of the nanowire) reduces the set to two. 

For the states at the edges of the nanowire subbands $k_z = 0$, 
one may obtain the following dispersion equation~\cite{Goupalov11}:
\begin{equation}
	\label{eq:disp}
	g J_{m^+}(kR) I_{m^-}(\kappa R) - G J_{m^-}(kR) I_{m^+}(\kappa R) = 0 \:,
\end{equation}
where $J_{m^{\pm}}(kR)$, $I_{m^{\pm}}(\kappa R)$ are 
respectively Bessel and modified Bessel functions of the integer index $m^{\pm} = m \pm \frac{1}{2}$,
$R$ is the NW radius,
\begin{equation}
	\label{eq:gG}
	\begin{split}
		g & = \frac{P k}{\alpha_v k^2 + E + E_g/2} \:, \\
		G & = \frac{P \kappa}{\alpha_v \kappa^2 - E - E_g/2}\:, \\
	\end{split}
\end{equation}
and
\begin{equation} 
	\label{eq:kk}
	\begin{split}
		k &= \sqrt{ \Xi + \Lambda } \:,\;\; \kappa = \sqrt{ \Xi - \Lambda }\:, \\
		\Xi &= \sqrt{\Lambda^2 + \frac{(4E^2 - E_g^2)}{4 \alpha_v \alpha_c}}\:, \\
		\Lambda &= \frac{E(\alpha_v - \alpha_c) - P^2 - (\alpha_v + \alpha_c)E_g/2}{2 \alpha_v \alpha_c}\:.
	\end{split}
\end{equation}

For a given $m$, we enumerate all roots of the dispersion Equation~\eqref{eq:disp}, $E_{m, n}$, with the index $n$ in a way that a positive $n$ refers to the $n$'th positive root and a negative $n$ refers to the $|n|$'th negative root. For each energy, there exist two eigenfunctions of the Hamiltonian~\eqref{eq:H},
which we denote as $\ket{\nu=0 \, \uparrow \, m \, n}$ and $\ket{\nu=0 \, \downarrow \, m \, n}$. Below, for the sake of brevity, we will sometimes omit one or two of these indices, where it cannot lead to a confusion, but the order of the indices will always be preserved. 
The explicit form of the eigenfunctions is
\begin{equation}
	\label{eq:psi_up_down}
	\begin{split}
		\ket{\uparrow m} & = \vec{\mathcal{E}}_0  \: \hat{\mu}_{\uparrow \, m} \:, \\
		\ket{\downarrow m} & = \vec{\mathcal{E}}_0 \: \hat{\mu}_{\downarrow \, m} 	\:,
	\end{split}
\end{equation}
where $\vec{\mathcal{E}}_0$ is the row vector of the basis functions \eqref{eq:0_basis} and $\hat{\mu}_{\uparrow(\downarrow) \, m}$ are envelope column bispinors
\begin{equation} \label{eq:bispinors}
  \hat{\mu}_{\uparrow m} =
	\begin{pmatrix}
		u(\rho) e^{i m^- \varphi}  \\
		0 \\

		0 \\
		i v(\rho) e^{i m^+ \varphi} \\
	\end{pmatrix}
	\:,\;
	\hat{\mu}_{\downarrow m} =
    \begin{pmatrix}
		0 \\
		   u(\rho) e^{- i m^- \varphi} \\
		i  v(\rho) e^{- i m^+ \varphi} \\
		0 \\
	\end{pmatrix} \:,
\end{equation}
The functions $u(\rho)$, $v(\rho)$ depend only on $\rho$ and take the form
\begin{equation}
	\label{eq:uv}
	\begin{split}
		u(\rho) &= N_m \left( J_{m^-}(k\rho) + c \, I_{m^-}(\kappa \rho) \right) \:, \\
		v(\rho) &= N_m \left( g J_{m^+}(k\rho) + c \, G I_{m^+}(\kappa \rho)  \right) \:,
	\end{split}
\end{equation}
where $N_m$ is the normalization constant, the constants $g$ and $G$ are defined in Eq.~\eqref{eq:gG} and the constant $c$ is
\begin{equation}
	\label{eq:c}
	c = - \frac{J_{m^-}(k R)}{I_{m^-}(\kappa R)} \:.
\end{equation}

The functions Eq.~\eqref{eq:psi_up_down}, as it was mentioned before, are eigenfunctions of the $\hat{J}_z$ with the following eigenvalues
\begin{equation}
	\label{eq:Jz_psi}
	\begin{array}{c}
		\hat{J}_z \ket{\uparrow \phantom{-} m} = \phantom{-} m \ket{\uparrow \phantom{-} m} \:, \\
		\hat{J}_z \ket{\downarrow \phantom{-} m} = -m \ket{\downarrow \phantom{-} m} \:, \\
		\hat{J}_z \ket{\uparrow -m} = -m \ket{\uparrow -m} \:, \\
		\hat{J}_z \ket{\downarrow -m} =\phantom{-} m \ket{\downarrow -m} \:.
	\end{array}
\end{equation}

Thus, the electron states at the nanowire subband edges ($k_z=0$), originating from a single valley, are characterized by the main quantum number $n$, projection $m$ of the total angular momentum on the NW axis, and parity. Numerical solutions of Eq.~\eqref{eq:disp} for a few energies near the 
band gap are shown in Table~\ref{tb:KP_spectra}.

\begin{table}[b]
  \caption{Subband-edge energies and quantum numbers for the first few subbands in the conduction and valence bands of a nanowire of the diameter $D=29.92$~\AA\mbox{} calculated in the framework of the isotropic ${\bm k\cdot \bm p}$ approximation with anisotropic parameters, listed in Sec.~\ref{sec:bulk_KP}. Symmetries of the \KP~states refer to the NWs of the $D_{3d}$ symmetry group and show distribution of the irreducible representations, $S$, associated with the states at the extrema of the corresponding energy subbands, over the longitudinal, $L_0$, and inclined, $L_{123}$ valleys.
} 
\label{tb:KP_spectra}
	\centering{
\begin{ruledtabular}
\begin{tabular}{d c d c d d}
\multicolumn{1}{c}{$E_{L_0}$, eV} & $S_{L_0}$ & \multicolumn{1}{c}{$E_{L_{123}}$, eV} & $S_{L_{123}}$ & n & m \\
\hline\hline
 3.45 & $\Gamma_{4}^- $ &  2.55 & $\Gamma_{5,6}^- \oplus 2\Gamma_4^-$ &  2 &  \frac{1}{2} \\
 3.04 & $\Gamma_{5,6}^-$ &  2.27 & $\Gamma_{5,6}^- \oplus 2\Gamma_4^-$ &  1 & -\frac{3}{2} \\
 2.99 & $\Gamma_{4}^- $ &  2.21 & $\Gamma_{5,6}^- \oplus 2\Gamma_4^-$ &  1 &  \frac{5}{2} \\
\hline
 1.81 & $\Gamma_{4}^+ $ &  1.39 & $\Gamma_{5,6}^+ \oplus 2\Gamma_4^+$ &  1 & -\frac{1}{2} \\
 1.77 & $\Gamma_{5,6}^+$ &  1.34 & $\Gamma_{5,6}^+ \oplus 2\Gamma_4^+$ &  1 &  \frac{3}{2} \\
\hline
 0.81 & $\Gamma_{4}^- $ &  0.65 & $\Gamma_{5,6}^- \oplus 2\Gamma_4^-$ &  1 &  \frac{1}{2} \\
\hline
-0.94 & $\Gamma_{4}^+ $ & -0.71 & $\Gamma_{5,6}^+ \oplus 2\Gamma_4^+$ & -1 & -\frac{1}{2} \\
\hline
-2.08 & $\Gamma_{5,6}^-$ & -1.48 & $\Gamma_{5,6}^- \oplus 2\Gamma_4^-$ & -1 & -\frac{3}{2} \\
-2.12 & $\Gamma_{4}^- $ & -1.53 & $\Gamma_{5,6}^- \oplus 2\Gamma_4^-$ & -1 &  \frac{1}{2} \\
\hline
-3.55 & $\Gamma_{4}^+ $ & -2.46 & $\Gamma_{5,6}^+ \oplus 2\Gamma_4^+$ & -1 & -\frac{5}{2} \\
-3.59 & $\Gamma_{5,6}^+$ & -2.51 & $\Gamma_{5,6}^+ \oplus 2\Gamma_4^+$ & -1 &  \frac{3}{2} \\
-4.09 & $\Gamma_{4}^+ $ & -2.83 & $\Gamma_{5,6}^+ \oplus 2\Gamma_4^+$ & -2 & -\frac{1}{2}
\end{tabular}
\end{ruledtabular}
	} 
\end{table}

\subsubsection{Parity-, $T$-, and $C$-symmetry of the states}
\label{subsubsec:P_and_T_and_C_symm}

The parity operator $\hat{\mathbf{P}}$ in cylindrical coordinates is defined as
\begin{equation}
	\label{eq:P_op}
	\hat{\mathbf{P}}f(\rho, \phi, z) = f(\rho, \phi + \pi, -z).
\end{equation}
Taking into account the parity of the Bloch functions we can easily establish the parity of the eigenstates Eqs.~\eqref{eq:psi_up_down}
\begin{equation}
	\label{eq:parities}
	\hat{\mathbf{P}} \ket{\uparrow(\downarrow) \, m} = (-1)^{m^+} \ket{\uparrow(\downarrow) \, m} \:.
\end{equation}
Note that parities of the functions $e^{i m^{\mp} \varphi}$ are different, but taking into account the parity of the Bloch functions,
\begin{equation}
	\label{eq:P_Bloch}
	\hat{P}\left(\ket{L_{0}^c}, \ket{L_{0}^v}\right) = \left(\ket{L_{0}^c}, \ket{L_{0}^v}\right) 
	\begin{pmatrix}
		-1 & 0 \\
		0 & 1
	\end{pmatrix}\:,
\end{equation}
it results in the overall parity of the functions $\ket{\uparrow(\downarrow) \, m}$ given by
Eq.~\eqref{eq:parities}.

Without magnetic field, the \KP\ Hamiltonian commutes with the time reversal operator 
$\hat{T}$. In the basis \eqref{eq:0_basis} it takes the
form $\left(I_{2} \otimes -i\sigma_{y}\right) \hat{K}_{0}$. Here $\hat{K}_{0}$ is the complex conjugation operator 
and $I_2$ is the $2\times2$ unit matrix.
Under the time reversal the eigenfunctions \eqref{eq:psi_up_down} transform one into another as
\begin{equation} 
	\label{eq:T_coupling}
	\begin{array}{l}
		\hat{T} \ket{\uparrow m} =  \hspace{7.78pt}\, \ket{\downarrow m} \:, \\
		\hat{T} \ket{\downarrow m} = - \ket{\uparrow m} \:,
	\end{array}
\end{equation}
which can be rewritten in a matrix form
\begin{equation} 
	\label{eq:T_coupling_matrix}
	\hat{T} \left( \ket{\uparrow m}, \ket{\downarrow m}\right) = \left( \ket{\uparrow m}, \ket{\downarrow m}\right)
	\begin{pmatrix}
		0 & -1 \\
		1 & 0
	\end{pmatrix}
	\:.
\end{equation}

Next we consider the charge conjugation operator \cite{Landau4_book} $\hat{C} = \gamma^{2} \hat{K}_{0}$, where $\gamma^{\mu}$ are Dirac matrices \cite{BjorkenDrell_book}, which act on the bispinors Eq.~\eqref{eq:bispinors}.
The Hamiltonian~\eqref{eq:H} is not invariant under the C-symmetry if $\alpha_{c} \ne \alpha_{v}$.
The charge conjugated Hamiltonian is
\begin{equation}
  H_{C} = -\gamma^{2}H^*\gamma^{2} \,,
\end{equation}
and the charge conjugation acting on the eigenfunctions produces  
functions which do not satisfy the dispersion equations. 
However, if $\alpha_{c} = \alpha_{v} $ then charge conjugation reverses the 
sign of the Hamiltonian 
\begin{equation}
  H_{C} = - H \,,
\end{equation}
thus exchanging conduction and valence bands.
In this case the main quantum number $n$ changes its sign and 
the wave functions transform as 
\begin{equation} \label{eq:C_coupling}
  \begin{split}
    \hat{C} \ket{\uparrow \hspace{-2pt} (\downarrow) \: m, n}  & = \text{sign}(n)(-1)^{m^-} \ket{\uparrow(\downarrow) -m, -n}.
  \end{split}
\end{equation}

We emphasize the fact that only in the case of a C-symmetric Hamiltonian 
the identity $E_{m, n} = -E_{-m, -n}$ holds.
In contrast to the time-reversal operator $\hat{T}$, the charge conjugation operator $\hat{C}$
exchanges the states with the opposite parity. 
Even though the charge symmetry does not hold in real lead chalcogenides, the 
corresponding analysis provides an insight into the structure of the electron energy levels.
If one starts from the Hamiltonian with $\alpha_{c} = \alpha_{v}$ then the states 
can be classified using the charge conjugation operator. 
An adjustment of the parameters $\alpha_{c}, \alpha_{v}$ to their actual values  
breaks the C-symmetry, but this does not affect the degeneracies of the levels and only slightly changes their energy positions.

This analysis is helpful in revealing the relations 
between the solutions with positive and negative energies. 
In particular, it guarantees that negative energy solutions have the 
opposite parities with respect to the corresponding positive energy solutions.

\subsubsection{Point group symmetry of the states originating from the longitudinal valley}
\label{subsubsec:point_symm}
In the previous subsection we characterized nanowire electron states, originating from a single valley of a bulk semiconductor,
by the main quantum number $n$, projection $m$ of the total angular momentum on the NW axis, and parity. This characterization assumes that the nanowire has an idealized shape of a circular cylinder, which implies the symmetry group $D_{\infty h}$, and does not take into account the fact that the symmetry of an actual nanowire is lower due to the underlying crystal structure. However, if one formally restricts the symmetry operations of the idealized nanowire by the symmetry operations of the actual nanowire point group, then one can assign an irreducible representation of this group to each electron energy level and thereby classify the states by their symmetry. In this paragraph we use this approach to analyze the symmetry of the states~\eqref{eq:psi_up_down}. It corresponds to the symmetry of the states originating from the  longitudinal valley of a [111] grown nanowire which is decoupled from the states originating from all the other valleys.

The highest possible point group of an actual NW
is $D_{3d}$ (cf. Fig.~\ref{fig:w_cells_BZ}). 
The point group $D_{3d}$ has six spinor representations $\Gamma_{4}^{\pm}$, $\Gamma_{5}^{\pm}$ 
and $\Gamma_{6}^{\pm}$ \cite{Koster}.
Representations $\Gamma_{5}^{\pm}$ and $\Gamma_{6}^{\pm}$ are
conjugated.
Therefore, the electron states of a nanowire can transform either under $\Gamma_{5}^{\pm} \oplus \Gamma_{6}^{\pm}$ (below we denote them as $\Gamma_{5,6}^{\pm}$) or under $\Gamma_4^{\pm}$. In systems, where the $z$ axis coincides with the $C_3$ symmetry axis, projection $m$ of the angular momentum $\hat{J}_z$ is defined modulo $3$. On the other hand, in $D_{3d}$ the states $\{\ket{\uparrow\, m},~\ket{\downarrow\, m} \}$ \eqref{eq:psi_up_down} with $|m|=3/2$ transform under $\Gamma_{5,6}^{\pm}$ while the states with $|m|=1/2$ transform under $\Gamma_{4}^{\pm}$~\cite{Koster}. 
This results in a simple criterion to classify the states in the $L_0$ valley in accordance with irreducible representations of the $D_{3d}$ group:
\begin{equation}
	\label{eq:criterion_D3d}
	\begin{array}{c | c}
		\Gamma_{5,6}^{P} & 2m~\text{mod}~3 = 0 \\
		\Gamma_4^{P} & 2m~\text{mod}~3 \ne 0
	\end{array},
\end{equation}
where $P$ is the parity of states, see Eq.~\eqref{eq:parities}.
One can verify that $C_2$ rotations around the axes $x$, $x'$, $x''$,
defined as
\begin{equation}
	\label{eq:x_ax_def}
\vec{x} \parallel [\bar{1}10] ,\;
\vec{x}' \parallel [0\bar{1}1] ,\;
\vec{x}'' \parallel [10\bar{1}]\:,
\end{equation}
transform the states $\{ \ket{\uparrow \, m},~\ket{\downarrow \, m} \}$ one into another. 
This proves that these pairs of functions form the bases of irreducible representations of 
the group $D_{3d}$. Note that the criterion Eq.~\eqref{eq:criterion_D3d} is still valid in $D_3$ NWs, except for the parity, since the group $D_{3d}$ may be represented as $D_3 \otimes C_i$.

In case of $C_{2h}$ NWs, there are four spinor representations. 
Two representations $\Gamma_3^{\pm}$ are conjugated to $\Gamma_4^{\pm}$ and the states 
transform either under $\Gamma_{3,4}^+ \equiv \Gamma_{3}^+\oplus\Gamma_{4}^+ $ or $\Gamma_{3,4}^-$.
Parity analysis requires microscopic consideration and is discussed in the next Section.

\subsection{Effective mass model for combinations of valley states}
\label{subsec:bloch_combinations}

In this section we extend the effective mass model to account for 
the valley structure of the states and classify the \KP\ solutions in accordance with 
the irreducible representations of the NW point group.
In order to do that we introduce bases in form of Eq.~\eqref{eq:0_basis} in each $L_{\nu}$, $(\nu=0,1,2,3)$ valley
\begin{equation}
	\label{eq:nu_basis}
	\vec{\mathcal{E}}_{\nu} = 
	\left(\ket{L_{\nu}^c}\ket{\uparrow},~\ket{L_{\nu}^c}\ket{\downarrow},~\ket{L_{\nu}^v}\ket{\uparrow},~ \ket{L_{\nu}^v}\ket{\downarrow}\right)
	\:
\end{equation}
and extend the set of solutions Eq.~\eqref{eq:psi_up_down} of the isotropic \KP~model by taking the valley index $\nu$ into account:
\begin{equation}
	\label{eq:psi_up_down_nu}
	\begin{split}
		\ket{\nu \uparrow m} & = \vec{\mathcal{E}}_{\nu} \: \hat{\mu}_{\uparrow \, m} \:, \\
		\ket{\nu \downarrow m} & = \vec{\mathcal{E}}_{\nu} \: \hat{\mu}_{\downarrow \, m} 	\:.
	\end{split}
\end{equation}

To analyse the states Eq.~\eqref{eq:psi_up_down_nu} let us consider the wave vectors $\vec{k}_{\nu}$, $(\nu = 0,1,2,3)$ of the $L$ valleys Eq.~\eqref{eq:valley_def}. In the bulk point group $O_h$ they form an irreducible star $S_{L_{0123}} = \{ \vec{k}_{\nu} \}_{\nu=0}^3$, but lowering the symmetry to the point groups $D_{3d}^{[111]}$ or $D_3^{[111]} \subset O_h$ ($C_3$ axis points along the [111] direction) one can find that the star breaks into irreducible ones $S_{L_0}$ and $S_{L_{123}}$
\begin{equation}
	\label{eq:S0S123}
	\begin{split}
		S_{L_0} & = \{ \vec{k}_{0} \} \:, \\
		S_{L_{123}} & = \{ \vec{k}_{1}, \vec{k}_{2}, \vec{k}_{3} \} \:.
	\end{split}
\end{equation}
In the $C_{2h}^{[111]} \subset D_{3d}^{[111]}$ the star $S_{L_{0123}}$ further breaks into irreducible stars $S_{L_2}$ and $S_{L_{13}}$,
\begin{equation}
	\label{eq:S2S13}
	\begin{split}
		S_{L_2} & = \{ \vec{k}_{2} \} \:, \\
		S_{L_{13}} & = \{ \vec{k}_{1}, \vec{k}_{3} \} \:.
	\end{split}
\end{equation}
It may be shown that within an irreducible star the valley index $\nu$ transforms as permutations
of the wave vectors in the star. This allows us to symmetrize the basis functions,
Eq.~\eqref{eq:nu_basis}, within the irreducible stars $S_{L_0}$, $S_{L_{123}}$ 
for $D_{3d}$ and $D_3$ NWs and within $S_{L_0}$, $S_{L_2}$, $S_{L_{13}}$ for $C_{2h}$ NWs.

Below we perform the symmetrization procedure only for the angular momentum $m=1/2$, \textit{i.e.} for the first confined electron states in each 
valley and we consider NWs with the groups $D_{3d}$ and $D_{3}$ together.

\subsubsection{Valley structure of states in $D_{3d}$ and $D_3$ NWs}

We have already shown in Sec.~\ref{subsubsec:point_symm} that the states originating
from the longitudinal valley ($S_{L_0}$ star) in $D_{3d}$ and $D_3$ NWs form the 
irreducible representation $\Gamma_{4}^{(-)}$, thus we consider the star $S_{L_{123}}$ here.

The matrices of the wave vector permutations for the $C_{3z}$ and $C_{2x}$ 
rotations, $T_{C_{3z}}$ and $T_{C_{2x}}$, within the star $S_{L_{123}}$ are 

\begin{equation}
	\label{eq:Bloch_transforms}
	T_{C_{3z}} = 
	\begin{pmatrix}
		0 & 0 & 1 \\
		1 & 0 & 0 \\
		0 & 1 & 0 
	\end{pmatrix}
	,
	~~~
	T_{C_{2x}} = 
	\begin{pmatrix}
		1 & 0 & 0 \\
		0 & 0 & 1 \\
		0 & 1 & 0 
	\end{pmatrix}
	\:.
\end{equation}
One may note that the matrices $T_{C_{3z}}$ and $T_{C_{2x}}$ together with the unit $3 \times 3$ 
matrix $\hat{U}_3$ form the $\Gamma_1 \oplus \Gamma_3$ representation of the group $D_3$.
Taking the spin into account, we can get a representation for the irreducible star $S_{L_{123}}$:
\begin{multline}
	\label{eq:D3reprS123}
	\left( \Gamma_1 \oplus \Gamma_3 \right) \otimes \Gamma_4 = 
	\left( \Gamma_1 \oplus \Gamma_3 \right) \otimes \Gamma_{5,6} = 
	\Gamma_{5,6} \oplus 2\Gamma_4
	\:.
\end{multline}
Taking into account the parity $P$ of the states \eqref{eq:parities} we get the full 
representation for the same star in the $D_{3d}$ NWs, which is just $\Gamma_{5,6}^P \oplus 2\Gamma_4^P$.

The states $\{ \ket{0 \uparrow 1/2}, \ket{0 \downarrow 1/2} \}$ are chosen 
in such a way that their 
rotation matrices coincide with the corresponding spin rotation matrix
\begin{equation}
	\label{eq:D12}
	D_{\frac{1}{2}}(\vec{n}, \omega) = \cos\left(\frac{\omega}{2}\right) - i \vec{n}\vec{\sigma} \sin\left(\frac{\omega}{2}\right) \:,
\end{equation}
where $\vec{n}$ is the rotation axis direction and  $\omega$ is the rotation angle.
Taking into account the permutation matrices $T$, the full rotation matrices for the states $\ket{\nu \uparrow \hspace{-2pt} (\downarrow)}$ are Kronecker products of the matrices $T$ and $M$ 
\begin{equation}
	\label{eq:DgS123}
	D(g) = T(g) \otimes M(g) \:.
\end{equation}
The symmetrization procedure for $D_{3d}$ and $D_{3}$ NWs consists of a simultaneous diagonalization  
of the matrices $D_{C_{3z}}$ and $D_{C_{2x}}$.
This can be achieved using the unitary matrix $V_{L_{123}}$ via
$V^{\dag} D V$. This matrix transforms the states as
\begin{multline}
	\label{eq:V123_states}
	\left( \ket{\Gamma_4^{(-)} \uparrow\hspace{-2pt}(\downarrow)}, \ket{\Gamma_4^{(-)} \uparrow\hspace{-2pt}(\downarrow)}, \ket{\Gamma_5^{(-)}}, \ket{\Gamma_6^{(-)}}  \right)  \\
	= \left( \ket{1 \uparrow\hspace{-2pt}(\downarrow)}, \ket{2 \uparrow\hspace{-2pt}(\downarrow)}, \ket{3 \uparrow\hspace{-2pt}(\downarrow)} \right) V_{L_{123}}\:.
\end{multline}
The explicit form of the matrix $V_{L_{123}}$ is given below

\begin{equation}
	\label{eq:V123}
	V_{L_{123}} = 
\left(
\begin{array}{ c c | c c | r | c }
\frac{1}{\sqrt{3}} & 0                  & 0                          & \frac{i        }{\sqrt{3}} &-\frac{i        }{\sqrt{6}} & \frac{i        }{\sqrt{6}} \\
0                  & \frac{1}{\sqrt{3}} & \frac{i        }{\sqrt{3}} & 0                          & \frac{i        }{\sqrt{6}} & \frac{i        }{\sqrt{6}} \\
\frac{1}{\sqrt{3}} & 0                  & 0                          & \frac{i\omega^2}{\sqrt{3}} &-\frac{i\omega  }{\sqrt{6}} & \frac{i\omega  }{\sqrt{6}} \\
0                  & \frac{1}{\sqrt{3}} & \frac{i\omega  }{\sqrt{3}} & 0                          & \frac{i\omega^2}{\sqrt{6}} & \frac{i\omega^2}{\sqrt{6}} \\
\frac{1}{\sqrt{3}} & 0                  & 0                          & \frac{i\omega  }{\sqrt{3}} &-\frac{i\omega^2}{\sqrt{6}} & \frac{i\omega^2}{\sqrt{6}} \\
0                  & \frac{1}{\sqrt{3}} & \frac{i\omega^2}{\sqrt{3}} & 0                          & \frac{i\omega  }{\sqrt{6}} & \frac{i\omega  }{\sqrt{6}}
\end{array}
\right)\:,
\end{equation}
where $\omega = \exp\left( \frac{2 \pi i}{3} \right)$. The vertical lines separate invariant subspaces $\Gamma_4^{(-)},~\Gamma_4^{(-)},~\Gamma_5^{(-)}$, and $\Gamma_6^{(-)}$.
The phase choice guarantees that all the $\Gamma_4^-$ states transform as basis spinors $\left\{ \ket{\uparrow}, \ket{\downarrow} \right\}$, and $\Gamma_5^-, ~\Gamma_6^-$ are coupled by the time reversal. 
The structure of the matrix $V_{L_{123}}$ enables one to analyze the valley structure of the states of a given symmetry.
From \eqref{eq:V123} it follows that, for the $D_{3d}$ and $D_3$ NWs, 
all the states with $m=1/2$ originating from the inclined valleys are equally distributed over these valleys.

\subsubsection{Valley structure of states in $C_{2h}$ nanowires}

The states in the third NW type with the point group $C_{2h}$ is much easier to symmetrize than in the cases of the point groups $D_{3d}$ and $D_3$. The $C_{2h}$ point group contains only one non-trivial element: rotation around the $\bm{x}'$ axis $C_{2x'}$, which permutes wave vectors in the star $S_{L_{13}}$.

We start from the stars $S_{L_0}$ and $S_{L_2}$. The states belonging to these stars can be symmetrized by the same matrix $V_{L_0}$
\begin{equation}
	\label{eq:V02_states}
	\left( \ket{\Gamma_3^-}, \ket{\Gamma_4^-}  \right)
	= \left( \ket{0(2) \uparrow}, \ket{0(2) \downarrow} \right) V_{L_{0}}\:, 
\end{equation}
where the matrix $V_{L_0}$ is
\begin{equation}
	\label{eq:V0}
	V_{L_0} = 
	\frac{1}{\sqrt{2}}
	\begin{pmatrix}
		-\omega   & -\omega   \\
		 \omega^2 & -\omega^2
	\end{pmatrix}
	\:.
\end{equation}
The rest of the states originated from the star $S_{L_{13}}$ can be symmetrized as follows
\begin{multline}
	\label{eq:V13_states}
	\left( \ket{\Gamma_3^+}, \ket{\Gamma_4^+}, \ket{\Gamma_3^+}, \ket{\Gamma_4^+}  \right) = \\
	= \left( \ket{1 \uparrow}, \ket{1 \downarrow}, \ket{3 \uparrow}, \ket{3 \downarrow} \right) V_{L_{13}}\: 
\end{multline}
with the matrix
\begin{equation}
	\label{eq:V13}
	V_{L_{13}} = 
\frac{1}{2}
\begin{pmatrix}
-\omega^2 & \omega^2 & \omega^2 & \omega^2 \\
  -\omega &  -\omega &  -\omega &   \omega \\
        1 &       -1 &        1 &        1 \\
        1 &        1 &       -1 &        1
\end{pmatrix}
	\:.
\end{equation}
One can see that the parity of the states at the $S_{L_{13}}$ star is opposite 
to the parity of the states at the $S_{L_0}$ and $S_{L_2}$ stars. The difference in parities arises from the 
parities of the atomic orbitals and the parity of the Bloch phase multiplier $e^{i \vec{k} \vec{r}}$. 
In the case of the $C_{2h}$ NWs, the inversion center is chosen at the 
point $a(\frac{1}{2},\frac{1}{4},\frac{1}{4})$ in the crystallographic 
coordinates (see Fig.~\ref{fig:w_cells_BZ}), which changes the parities of the
functions $\ket{L_{1,3}^{c,v}}$ to the opposite, as compared to the case when the
inversion center is chosen at the $(0,0,0)$ cation (the same as in bulk crystal),
while keeping parities of the Bloch functions $\ket{L_{0,2}^{c,v}}$ the same. 


\section{Tight-Binding method} \label{sec:TB}

In this Section we discuss application of the empirical tight-binding method to PbSe nanowires.

The tight-binding method is based on the expansion of the electron wave function $  \Phi^s (\vec{r})$
in the local basis 
of atomic-like orbitals $\phi_{\sigma}(\vec{r})$,
which are assumed to be orthogonal \cite{Lowdin50}:
\begin{equation}\label{eq:TB_psi_r}
  \Phi^s (\vec{r}) = \sum_{n \sigma} C_{n \sigma}^s \phi_{\sigma}(\vec{r} - \vec{r}_{n} )\:,
\end{equation}
where $n$ enumerates the atoms and the index $\sigma$ runs through different orbitals. 
For the $sp^3d^5s^*$ variant of the 
method these are $s$, $s^*$, three $p$, and five $d$ orbitals multiplied by spin-up and spin-down basis spinors which results in the total of
twenty orbitals per atom.
In this basis,  the Schr\"odinger equation reduces to the eigenvalue problem for a
sparse matrix:
\begin{equation}\label{eq:TB_Ham_r}
  \sum_{n',\varsigma} H_{n\sigma,n'\varsigma} C^s_{n'\varsigma} = E_s C^s_{n\sigma} \,.
\end{equation}
Here, the eigenvalues $E_s$ correspond to the energies of the 
electron state $s$ and eigenvectors $C_{n\sigma}$ provide the coefficients in the 
expansion~\eqref{eq:TB_psi_r} for this state.
We use the nearest neighbor approximation, thus neglecting the matrix elements 
of the tight-binding Hamiltonian between the atoms that are not directly connected by 
a chemical bond.
The explicit form of the Hamiltonian matrix elements,
$H_{n\sigma,n'\varsigma}$,
may be found in \cite{Slater54} or \cite{Podolskiy04}.
The spin-orbit interaction is introduced following Chadi \cite{Chadi77}.
The tight-binding parameters~\cite{Poddubny12} are chosen to accurately reproduce experimental effective masses at the $L$ point as well as the electron energies at the high-symmetry points of the Brillouin zone which were calculated in Ref.~\cite{Svane10} using the $GW$-approximation.

For a nanowire, we introduce the wave vector $K_z$ along the 
growth direction. Its relation to the wave vector $k_z$ of Sec.~\ref{sec:kp} 
will become apparent later, in Sec.~\ref{subsec:TB_Fourier}. The tight-binding 
wave function for a NW takes the form
\begin{equation}\label{eq:TB_psi}
  \Phi_{K_z}^s (\vec{r}) = \sum_{n \sigma} e^{i  K_z z_n } \, C_{n \sigma}^s (K_z) \, \phi_{\sigma}(\vec{r} - \vec{r}_{n} )\:,
\end{equation}
where $s$ is the band index, the index $n$ runs over atoms in the elementary cell of the nanowire, 
and $z_n$ is the $z$-coordinate of the $n$-th atom.
Therefore, the tight-binding Hamiltonian for a nanowire is a finite matrix which depends on the wave vector $K_{z}$ and satisfies the equation
\begin{equation}\label{eq:TB_Ham}
  \sum_{n',\varsigma} H_{n\sigma,n'\varsigma}(K_z) C^s_{n'\varsigma}(K_z) = E_s(K_z) C^s_{n\sigma}(K_z) \,.
\end{equation}
Its matrix elements are related to the matrix elements of the bulk Hamiltonian via 
$H_{n\sigma,n'\varsigma}(K_z) = e^{i K_z \delta z_{nn'} }H_{n\sigma,n'\varsigma}$, where 
$\delta z_{nn'}$ is the $z$-component of the chemical bond vector between atoms $n$ and $n'$.


The nanowire structures used in our tight-binding calculations can be conceived as being carved out from a bulk PbSe crystal and inscribed in a circular cylinder with the axis parallel to the [111] direction and passing through 
(a) an atomic site, (b) the point with the 
coordinates $\frac{a}4(1,1,1)$, or (c) the midpoint of a chemical bond, as shown in Fig.~\ref{fig:w_cells_BZ}.
We do not use surface passivation in our  calculations. This is justified by
the strong ionicity of the  chemical bonds in the lead chalcogenides, that dramatically reduces the surface impact on the confined states~\cite{Delerue2004b,Abhijeet11,Poddubny12} as compared to 
covalent semiconductors such as Si~\cite{Delerue2004}.
In realistic nanowires such passivation might be necessary to compensate for the unbalanced surface charge~\cite{Leitsmann2009}, however, its impact on the energy spectrum should be relatively weak.

\subsection{Point group symmetry}
\label{subsec:TB_symmetry}

In order to analyze the symmetry of the tight-binding wave function one should determine how the symmetry operations, in particular rotations, affect the coefficients $C_{n\sigma}^s(K_z)$.

First, a rotation causes the atoms to change their positions.
Application of 
the group element $g$ transforms the atom $n$ with the coordinates $\vec{r}_n$ 
into the atom $n'$ with the coordinates $\vec{r}_{n'}=g^{-1}\vec{r}_n$. 
This 
transformation is described by
the permutation matrix $P_{g}$ which accumulates 
information about rotation of the atoms inside the NW.
This matrix implicitly depends on the quasi-wave vector $K_z$, as far as the transformations of the coefficients $C_{n\sigma}^s(K_z)$ are concerned.

Second, a rotation also affects the atomic orbitals forming the basis of the tight-binding Hamiltonian.
Indeed, the $\pi/2$ rotation of the function $p_x$ around $z$ results 
in the function $p_y$ which should be reflected in the corresponding change of 
the tight-binding coefficients. Formally, this may be described as follows.
In the logic of the Slater-Koster notations~\cite{Slater54} the $s$, $p$, and $d$ 
functions transform as the basis functions of the momentum 0, 1, and 2, respectively. 
As a result, the action of the symmetry element $g$ on these functions can be 
represented by Wigner's D-matrices for these momenta. 
The tight-binding coefficients before and after a rotation operation $g$, are connected 
by the matrix $D_{orb}(g)$:
\begin{multline}
  D_{orb}(g) = D_{\frac{1}{2}}(g)  \\ \otimes T^{-1} 
  \begin{pmatrix}
   D_{0}^+(g) & & & \\
   & D_{1}^-(g) & & \\
   & & D_{2}^+(g) & \\
   & & & D_{0}^+(g) 
   \end{pmatrix} T\:,
\end{multline}
where the matrix $T$ transforms the spherical harmonics into 
the tesseral ones~\cite{Whittaker2012course} which are 
used in tight-binding calculations \cite{Slater54,Podolskiy04}.

The tight-binding coefficients before and after application of
the symmetry operation $g$ are connected by the matrix  $D_{TB}(g)$ which
is constructed as Kronecker's product of the permutation matrix 
$P_g$ and the matrix $D_{orb}(g)$:
\begin{equation}\label{eq:TB_rot}
  D_{TB}(g) = P_g \otimes D_{orb}(g) \,.
\end{equation}

In the basis where the tight-binding Hamiltonian, $H_{n\sigma,n'\varsigma}(K_z)$, is diagonal, 
the transformation matrix \eqref{eq:TB_rot} has a block-diagonal 
form, with the blocks corresponding to irreducible representations 
of the nanowire symmetry group. This allows us to 
associate the tight-binding wave functions with the irreducible 
representations of the NW symmetry group.
The results of this procedure are presented for the eight low-energy conduction-band states in Table~\ref{tb:TB_symm_GCSs}. These results are in agreement with the symmetry analysis of Sec.~\ref{subsec:bloch_combinations}.

\begin{table}[t]
  \caption{
  Symmetry assignment for the eight low-energy conduction-band states obtained in the tight-binding calculations for NWs of three different symmetry groups.}
  	\label{tb:TB_symm_GCSs}
\begin{ruledtabular}
\begin{tabular}{ c  c }
		\multicolumn{1}{c}{\hfill point group} & \multicolumn{1}{c}{representation}  \\ 
		\hline
    		$D_{3d}$ & $3\Gamma_4^- \oplus \Gamma_{5,6}^- $  \\ 
		$D_{3}$ & $3\Gamma_4 \oplus \Gamma_{5,6} $ \\ 
		$C_{2h}$ & $ 2 \Gamma_{3,4}^- \oplus 2 \Gamma_{3,4}^+ $ \\
  	\end{tabular}
\end{ruledtabular}
\end{table}

\subsection{Fourier analysis}
\label{subsec:TB_Fourier}

In order to analyze how various states of the nanowire are related to different valleys of the bulk PbSe in the framework of the tight-binding method,
we introduce the function
\begin{equation}
	\label{eq:TB_F_psi_K}
	F_{K_z \omega}^{s} (\vec{\kappa}) = \sum_{m} e^{ i ( K_z z_m - \vec{\kappa}\vec{r}_{m})} C_{m \omega}^s ({K_z})\:,
\end{equation}
where $s$ is the state index,
the composite index $\omega$ labels  both the sublattice (anions or cations) 
and the atomic orbitals, 
$m$ enumerates atoms from one of the sublattices (specified in $\omega$) within the elementary cell, $\vec{r}_m$ are the coordinates of these atoms, and $z_m$ is the $z$-component of $\vec{r}_m$.
If the radius of the nanowire tends to infinity, $R \rightarrow \infty$, then the function \eqref{eq:TB_F_psi_K} 
forms $\delta$-like peaks near the projections of valley extrema onto the (111) plane 
and equivalent points in the reciprocal space.
For NWs of finite radius, the width of the peaks depends on the NW radius as $\propto R^{-1}$.

The elementary cell of a nanowire contains six atomic layers in the planes perpendicular to the growth direction (cf. Fig.~\ref{fig:w_cells_BZ}). The $z$-coordinates of these layers, $z_m$, are $0$, $a/\sqrt3$, and $2a/\sqrt3$ for the cation layers and $a/2\sqrt3$, $\sqrt{3}a/2$, and $5a/2\sqrt3$ for the anion layers. 
Taking this into account and substituting
\begin{equation}
	\label{eq:kappa_sub}
	\vec{\kappa} = K_z \, \hat{\vec{z}} + \vec{\kappa}_{\rho} + n \, \vec{b}_k \:,
\end{equation}
where $\vec{\kappa}_{\rho} \perp \vec{z}$ and $n=0,1,2$,  
one can rewrite Eq.~\eqref{eq:TB_F_psi_K} as 
\begin{equation}
	\label{eq:TB_F_psi_K_sub}
	F_{K_z \omega}^{s} (\vec{\kappa}_{\rho}, n) = \sum_{m} e^{-i ( \vec{\kappa}_{\rho} + \vec{b}_k \, n)\vec{r}_{m}} C_{m \omega}^s (K_z)\:.
\end{equation}
Here, $K_z$ can assume any value within a one-dimensional Brillouin zone of the nanowire. 

We further
introduce the local density of states (LDOS) in the reciprocal space:
\begin{equation}
	\label{eq:k_ldos}
	n_{K_z}^S(\vec{\kappa}_{\rho}, n)  = \sum_{s\in S}\sum_{\omega} |F_{K_z \omega}^{s} (\vec{\kappa}_{\rho}, n)|^2 \:,
\end{equation}
where the index $S$ refers to the set of states within one degenerate energy level.
As in Sec.~\ref{subsec:bloch_combinations}, we will be primarily interested in the states near the
extrema of the nanowire energy subbands in the conduction and valence bands. 
Therefore, we set $K_z=k_0$ (cf. Fig.~\ref{fig:w_cells_BZ}d).
and distinguish the
three functions $n_{k_0}^S(\vec{\kappa}_{\rho}, 0)$, $n_{k_0}^S(\vec{\kappa}_{\rho}, 1)$ 
and $n_{k_0}^S(\vec{\kappa}_{\rho}, 2)$ corresponding to the three [111] 
planes in Fig.~\ref{fig:w_cells_BZ}d). We note that the quasi wave vectors $K_z$, used in this section, and $k_z$, used in Sec.~\ref{sec:kp}, are related by $k_z=K_z-k_0$. Therefore, the assignment of
$K_z=k_0$ corresponds to $k_z=0$.

As explained in Section~\ref{sec:wire_structure}, the wave vectors from the first Brillouin zone 
of the bulk crystal, parallel to the [111] direction, which fill the interval $(-k_0,k_0]$ 
in the nanowire should be mapped onto
the first Brillouin zone of the NW which is three times smaller: $(-b_k/2,b_k/2]$ (zone folding). 
The number $n$ in \eqref{eq:TB_F_psi_K_sub} allows for the unfolding of the band structure of the 
wire (see e.g. Ref.~\cite{Boykin16}) onto the band structure of the bulk crystal.
The use of all the three functions is superfluous, as they
are connected by the translations
\begin{multline}
	\label{eq:n_k_diff}
	n_{k_0}^S(\vec{\kappa}_{\rho}, 0) = 
        n_{k_0}^S(\vec{\kappa}_{\rho} - \vec{b}_1 + \vec{b}_k, 1) \\
        = n_{k_0}^S(\vec{\kappa}_{\rho} + \vec{b}_1 - \vec{b}_k, 2) \:.
\end{multline}
Therefore, one can use only one of the planes in the valley analysis.

\begin{figure}[tbp]
  \centering{\includegraphics[width=\linewidth]{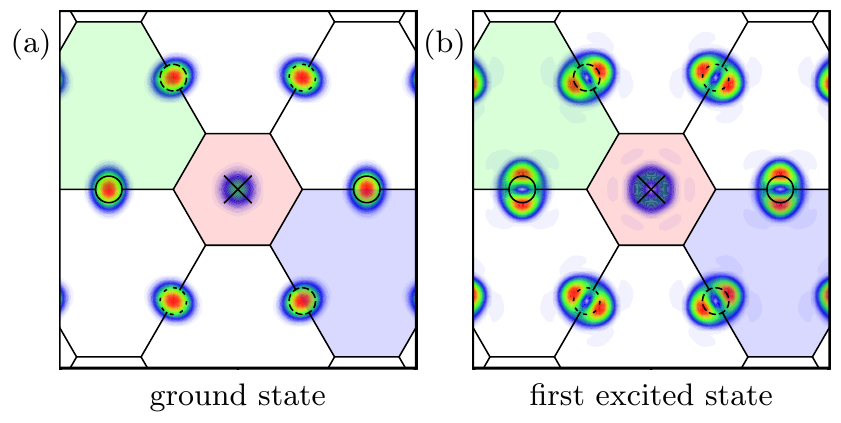}}	
	\caption{(Color online) LDOS in reciprocal space for the ground (a)  and the first excited (b) conduction state in the $D_{3d}$ NW with the $29.9$~\AA~diameter. 
	}
	\label{fig:confined_fourier}
\end{figure}
The analysis of the reciprocal space LDOS allows one
to distinguish states of the same symmetry but with different main quantum numbers \cite{Dohnalova2013,Hapala2013}.
The reciprocal space LDOS for two levels with similar valley structure, but 
different main quantum numbers is illustrated in Fig.~\ref{fig:confined_fourier}.  Each reciprocal space maximum for the ground electron state corresponds to two maxima for the first excited state.

\begin{figure*}[tbp]
  \centering{\includegraphics[width=1\linewidth]{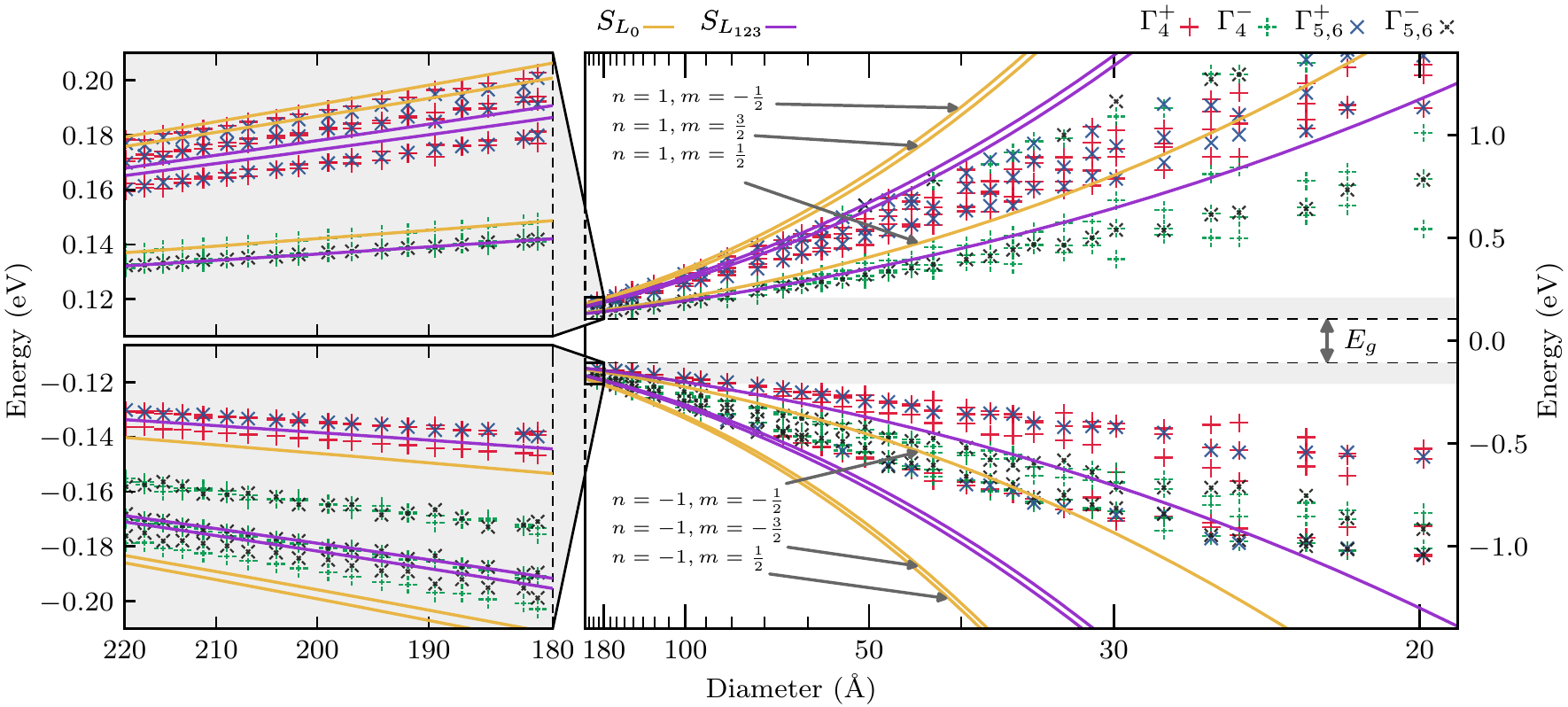}}	
	\caption{(Color online) Energy of the first 24 energy levels in the conduction and valence 
        bands in $D_{3d}$ nanowire as a function of the NW diameter. Tight-binding calculations are 
        represented as colored symbols ``+'''s and ``$\times$'''s, encoding irreducible 
        representations of states. 
        The results of \KP~calculations are shown by solid lines, purple for the inclined 
        states, and yellow for the longitudinal ones. $E_g$ is the band gap energy, 
        arrangement of the \KP~states is listed in Table~\ref{tb:KP_spectra}. 
        Left panel shows a blow up of the region corresponding to large NW diameters.
}	\label{fig:E-r-dependence}
\end{figure*}

\section{Results}
\label{sec:results}

We have performed tight-binding calculations of energy 
spectra for NWs grown along [111] and having different symmetries. We will first present our results for the states at the extrema of the nanowire subbands as functions of 
the nanowire diameter. 
We will limit our consideration by the 24 states in the conduction and 24 states in the valence band. 
These states correspond to the three multiplets in each of the bands most close to the nanowire Fermi level (cf. Fig.~\ref{fig:levels_scheme}).
Each of the multiplets originates from an eight-fold spin- and valley-degenerate level. These levels are first split into two-fold and six-fold degenerate  
levels due to the effective mass anisotropy. When the valley mixing at the 
nanowire surface is taken into account then each of the eight-fold degenerate levels is split into four doublets, see Fig.~\ref{fig:levels_scheme}.

\subsection{Valley structure}\label{subsec:valley_structure}

\subsubsection{\texorpdfstring{$D_{3d}$}{D3d} and \texorpdfstring{$D_3$}{D3} NWs}

\begin{figure}[tbp]
\vspace{-0.25cm}
  \centering{\includegraphics[width=1\linewidth]{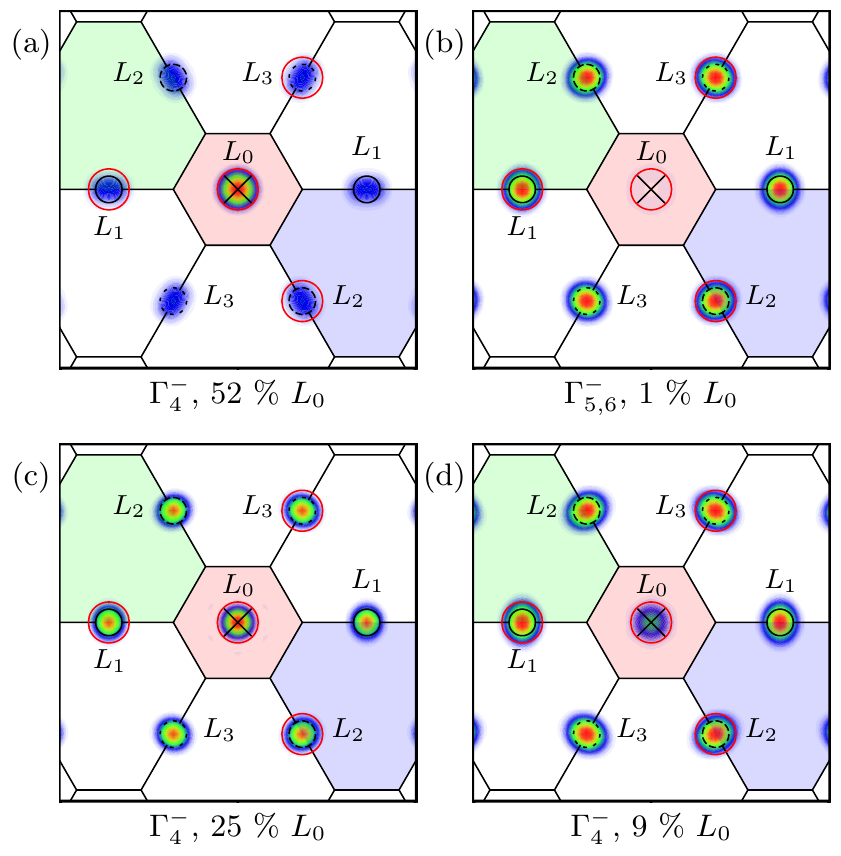}}	
\caption{(Color online) Reciprocal space LDOS for four 
  lowest-energy doublets of states in  the conduction 
  band of $D_{3d}$ nanowire with the 29.9~\AA~diameter. 
  The positions of the longitudinal valley $L_0$ are 
  indicated by the cross
  and  black circles 
  (solid, dashed, dotted) show positions of the inclined valleys 
  ($L_1, L_2, L_3$), see Sec.~\ref{subsec:TB_Fourier}.
  We also give the estimation (in percent) of $L_0$ valley contribution obtained 
  by integration of the LDOS near the peaks (within the areas enclosed in red circles).
  }
\label{fig:Fourier_wire_D3d}
\end{figure}

First we discuss the $D_{3d}$ nanowires.
The states originating from the lowest-energy eight-fold degenerate multiplet in the conduction band 
(see Fig.~\ref{fig:levels_scheme}) form bases of $\Gamma_{5,6}^{-}$ and three $\Gamma_4^{-}$ irreducible representations. 
As discussed in Sec.~\ref{subsec:bloch_combinations}, all the $\Gamma_4^{-}$ pairs of states have different LDOS in the reciprocal space. One pair has significant amount of density near the longitudinal valley $L_0$ and the others have their densities mostly at the inclined valleys $L_1, L_2, L_3$ (see Fig.~\ref{fig:Fourier_wire_D3d}). Other multiplets are split into the states that transform according to the irreducible representations, see Table~\ref{tb:KP_spectra}.

The results of the calculations for six multiplets (three in the conduction and three in the valence band) are presented in Fig.~\ref{fig:E-r-dependence}.
States transforming according to different irreducible representations 
are shown by crosses of 
different colors. This allows one to separate unambiguously the 
states originating from the ${\bm k\cdot\bm p}$ levels with $n=1$, $m=1/2$ 
from the pair of levels $n=1$, $m=-1/2,3/2$ since these levels 
have different parities (see Table~\ref{tb:KP_spectra}). An additional analysis is needed to 
reveal the valley structure of the levels because there are several 
sublevels with the same symmetry in all multiplets.

One can see that the valley splitting can exceed the energy distance between different multiplets for nanowire diameters below $40$~\AA. For instance, from the \KP~theory one expects to find in the conduction band four odd-parity doublets followed by eight even-parity doublets (see Table~\ref{tb:KP_spectra}). However, for some diameters, the highest-energy states shown in Fig.~\ref{fig:E-r-dependence} have the odd parity and, therefore, originate from the odd-parity multiplets which correspond to the top three lines of
Table~\ref{tb:KP_spectra}. Thus, the inter-valley coupling is comparable to the 
quantum confinement energy.

The strong inter-valley coupling
makes it impossible to track the origin 
of the states from the simple comparison with the \KP\ approximation, especially 
for the excited states. 
In order to resolve the valley structure of the  ground state multiplet, we use the analysis 
of reciprocal space LDOS explained in Section~\ref{subsec:TB_Fourier}.

In Fig.~\ref{fig:Fourier_wire_D3d} we show the reciprocal space LDOS for 
the first four levels in the conduction band of a nanowire with the diameter 29.9~\AA.
In accordance with the symmetry analysis of Sec.~\ref{subsec:bloch_combinations}, 
the level with the symmetry $\Gamma_{5,6}^-$
shows almost 100\% fraction of the inclined valleys. This is 
expected because the corresponding states can mix only with the states of the same symmetry 
and there are no states of the same symmetry within the energy range $\sim 1$~eV.
There are also two levels of the symmetry $\Gamma_4^-$ which are combinations of the states  
of inclined valleys with a small admixture 
of the longitudinal valley $L_0$ and one level of the symmetry 
$\Gamma_4^-$ with a strong admixture of the state from the longitudinal valley. The decomposition of the confined states over different valleys is clearly seen in Fig.~\ref{fig:Fourier_wire_D3d}. 

\begin{figure}[tbp]
  \centering{\includegraphics[width=1\linewidth]{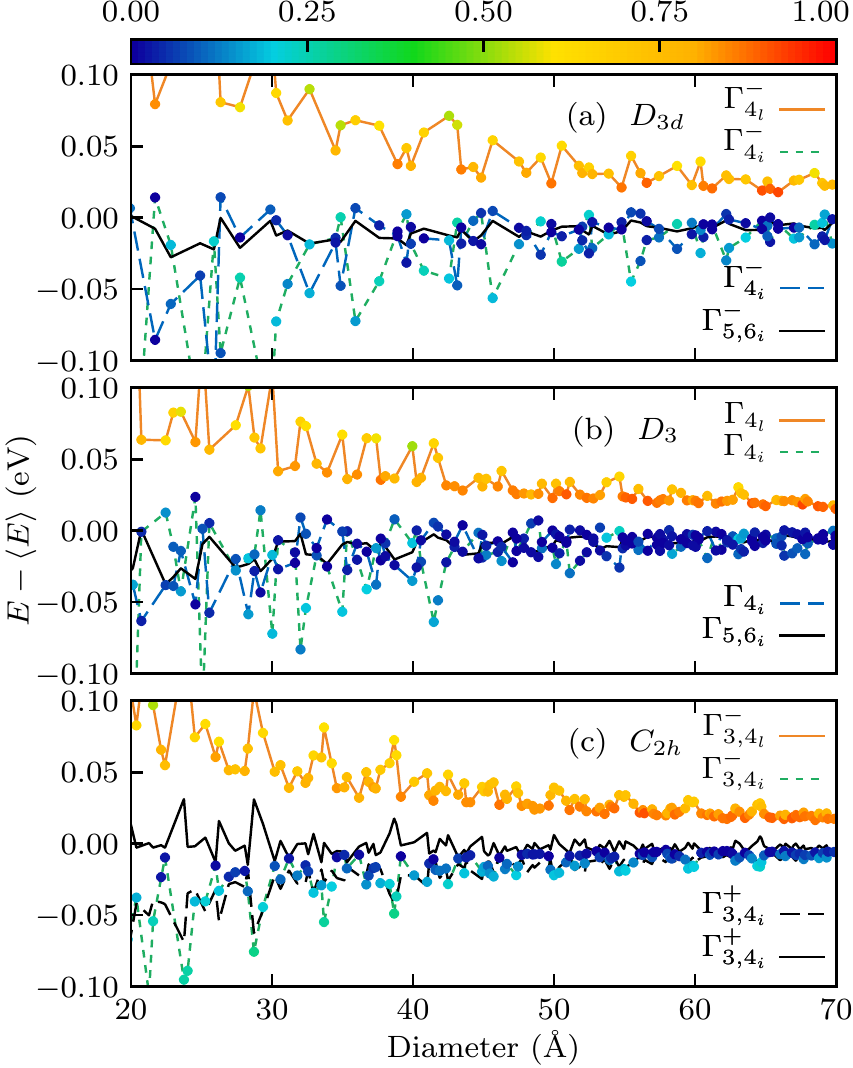}}	
\caption{
(Color online) Valley splittings for lowest conduction band states in $D_{3d}$ (a), $D_{3}$ (b) and $C_{2h}$ (c) NWs as 
a function of the NW diameter.
The colors of the dots encode the fraction of the longitudinal valley contribution $L_{0}$ to the confined state extracted from the $\bm k$-LDOS Eq.~\eqref{eq:k_ldos}.
}
 \label{fig:valley_splitting_C123}
\end{figure}

The reciprocal space LDOS analysis can be used to trace the dependence of the valley splittings on 
the NW diameter. This dependence is highly oscillating, which is explained by the fact that even a small variation in the NW radius results in substantially different arrangements of the surface atoms, similar to the quantum dot case~\cite{friesen2010,Poddubny12}. In Fig.~\ref{fig:valley_splitting_C123}(a)--Fig.~\ref{fig:valley_splitting_C123}(c)
we show only the first four double degenerate states in the conduction band
for the NW diameters $16\div 70$~\AA\mbox{} for $D_{3d}$ (a),  $D_{3}$ (b) and  $C_{2h}$ (c) NWs.  
The energies in Fig.~\ref{fig:valley_splitting_C123} are counted from the mean energy of the multiplet and are shown as functions of the nanowire diameter. The color of the points in Fig.~\ref{fig:valley_splitting_C123} encodes the contributions of the longitudinal ($L_0$) and inclined ($L_{1,2,3}$) valleys to a given energy level. The red (blue) color corresponds to the predominant contribution of the longitudinal (inclined) valleys.

It is clearly seen that the valley composition strongly correlates with 
the levels repulsion. 
The more pure are the states of the $\Gamma_{4}^-$ symmetry  originating from the longitudinal and inclined valleys, respectively, the lower is the splitting between these valleys in  Fig.~\ref{fig:valley_splitting_C123}(a). This correlation is somewhat obscured by the presence of the two levels originating from the inclined valleys. Below we will see that, for the NWs of the $C_{2h}$ symmetry, it is more pronounced.

Next we briefly discuss the  $D_3$ NWs, which have no inversion center. 
The dependences of energies of the confined states on the NW diameter are quite similar to these of the $D_{3d}$ case presented in Fig.~\ref{fig:E-r-dependence} and are not shown here.
The valley splittings,  shown in Fig.~\ref{fig:valley_splitting_C123}(b), oscillate with a somewhat smaller amplitude
than in the $D_{3d}$ case. This is probably due to a more symmetric surface of the $D_3$ NWs.
\subsubsection{\texorpdfstring{$C_{2h}$}{C2h} nanowires}
The $C_{2h}$ NWs are substantially different from the $D_{3d}$ and $D_3$ NWs. 
The group $C_{2h}$ has four spinor representations, $\Gamma_3^{\pm}$ and $\Gamma_4^{\pm}$, and all of 
them appear in the decomposition of the ground conduction (or valence) band state. The representations $\Gamma_3$ and $\Gamma_4$ are conjugate, and, therefore, the corresponding levels are doublets $\Gamma_{3,4}$. The fact that states of different parities originate from a single \KP\ level can be explained by  the change in the position of the inversion center as compared to a bulk crystal.  As shown in Sec.~\ref{subsec:bloch_combinations}, the odd-parity states are formed by the states from the $L_0$ and $L_2$ valleys, while the even-parity states are formed by the states from the $L_1$ and $L_3$ valleys. This result can be also obtained in the framework of the tight-binding method, see Sec.~\ref{subsec:TB_symmetry}.

\begin{figure}[tpb]
	\vspace{-0.25cm}
  \centering{\includegraphics[width=1\linewidth]{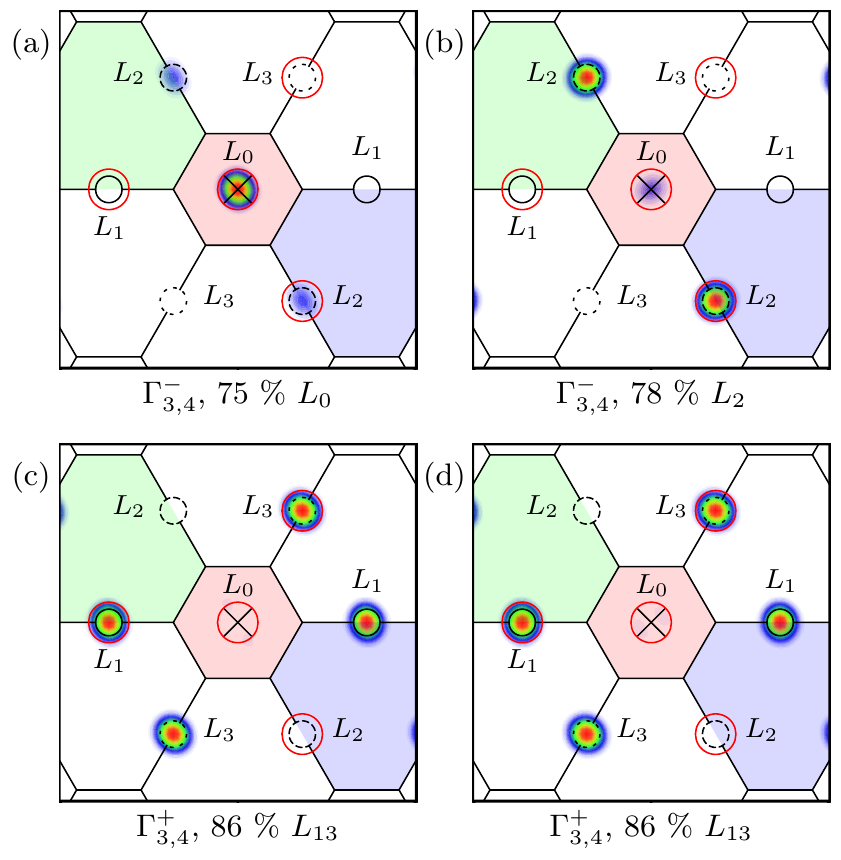}}	
	\caption{(Color online) Same as in Fig.~\ref{fig:Fourier_wire_D3d}, but for 
        $C_{2h}$ NW of diameter $D = 30.2$~\AA. 
	}
	\label{fig:Fourier_wire_C2h}
\end{figure}

In Fig.~\ref{fig:Fourier_wire_C2h} we show the reciprocal space LDOS for 
the first four levels in the conduction band of a nanowire with diameter 30.2~\AA.

Neglecting the valley mixing, the even-parity states $\Gamma_{3,4}^+$, originating from the valleys 
$L_1$ and $L_3$, are degenerate, while the odd-parity states $\Gamma_{3,4}^-$, originating 
from the $L_0$ and $L_2$ valleys, are split due to the valley
anisotropy. 
Thus, it is natural to expect that the states 
originating from the \KP\ level $n=1$, $m=1/2$ with 
the symmetry $\Gamma_{3,4}^+$ are evenly distributed between the two valleys while each of the split 
$\Gamma_{3,4}^-$ levels are predominantly contributed by the states from either 
$L_0$ or $L_2$ valley.
This qualitative analysis is in agreement with the calculated reciprocal space 
LDOS of the states presented in Fig.~\ref{fig:Fourier_wire_C2h}.
{The dependence of the first four levels in the conduction band
on the nanowire diameter is shown in Fig.~\ref{fig:valley_splitting_C123}(c).
For the odd-parity levels $\Gamma_{3,4}^-$, the valley composition strongly correlates with 
the levels repulsion. The larger is admixture of the $L_2$ valley to the state dominated by the $L_0$ valley (and vice versa), the stronger is the levels repulsion.
Both levels exhibit 
strong oscillations with an amplitude up to 100~meV for small radii and decreasing with the NW radius.  The strong correlation between the splitting of the levels and the degree of the valley admixture, encoded by the color in Fig.~\ref{fig:valley_splitting_C123}(c), clearly indicates that these oscillations are caused by the valley mixing at the interface.}

\begin{figure}[tpb]
  \centering{\includegraphics[width=0.7\linewidth]{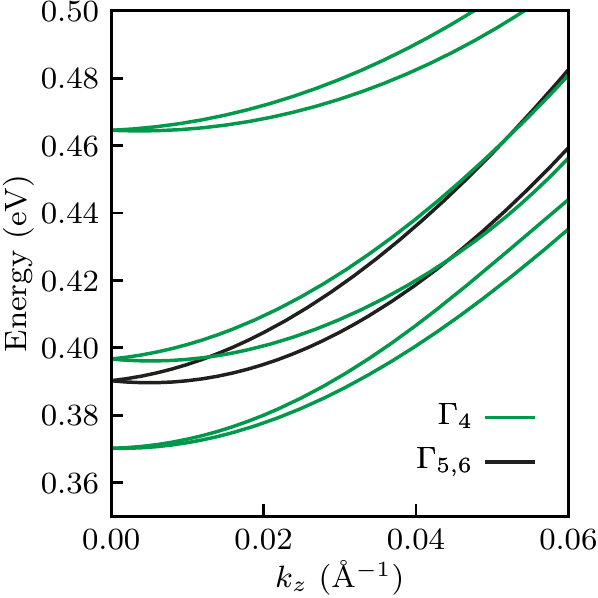}}
	\caption{(Color online) Spin splittings of the ground conduction states in the 
        $D_3$ NW with the $40$~\AA~diameter. 
        By green (black) lines we show $\Gamma_4$ ($\Gamma_{5,6}$) states.
        Wave vector $k_{z}$ is measured from the band edge.
	}
	\label{fig:spin_splitting}
\end{figure}

\subsection{Spin splitting}

In Sec.~\ref{subsec:valley_structure} we discussed the valley structure of different 
electron levels at the subband extrema for nanowires of different point groups. 
From the \KP\ analysis one could expect, that due to the spatial inversion 
in the bulk lead chalcogenides, it is impossible to obtain the fine spin structure 
of the energy levels in the nanostructures in the absence of magnetic field. 
We remind that the time-reversal symmetry together with the spacial 
inversion symmetry leads to the spin degeneracy of all levels
\begin{equation}\label{eq:Kramers}
E_{\uparrow(\downarrow) m,n}(K_z) = E_{\uparrow(\downarrow) m,n}(-K_z)\:.
\end{equation}

\begin{figure}[tpb]
  \centering{\includegraphics[width=\linewidth]{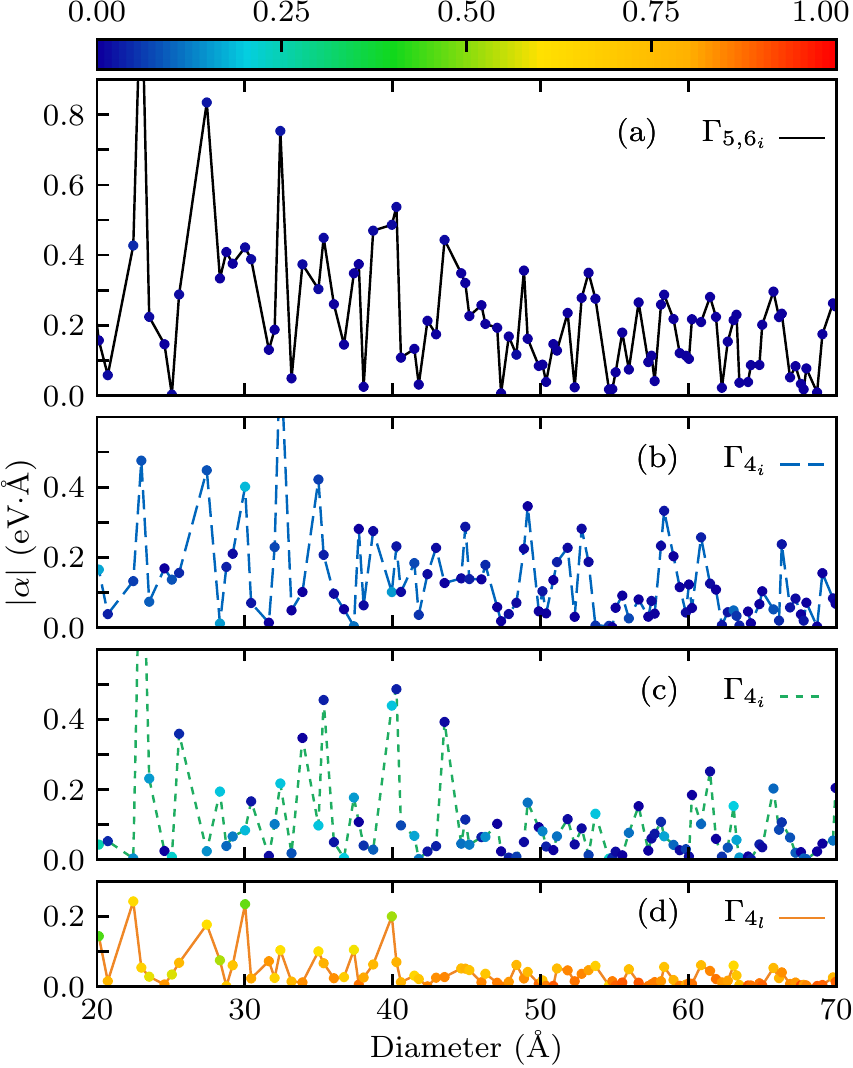}}
	\caption{(Color online) Spin splittings of four 
        conduction band doublets in NWs of $D_3$ symmetry 
        as a function of the NW diameter. 
        Color of the dots encodes the contribution of the longitudinal valley to the confined state.
	}
	\label{fig:spin_splitting_r}
\end{figure}

However, the atomistic texture near the surface may break the spacial inversion, 
even in centrosymmetric materials \cite{Nestoklon06,Nestoklon08}
and this is the case of the NWs with the point group $D_{3}$.
This group has no inversion center and  
Eq.~\eqref{eq:Kramers} no longer holds for $K_z \neq 0$. 
Therefore, spin-dependent splittings of the states become possible.
The spin splittings are linear in the wave vector $K_z$,
\begin{equation}
	\label{eq:spin_split}
	\Delta E_{\rm spin}^{S} = \alpha_{S} (K_{z}-k_{0})\equiv \alpha_{n}k_{z},
\end{equation}
where $\alpha_{S}$ is the splitting constant for the $S$'th couple of states. 
In Fig.~\ref{fig:spin_splitting} we show the energy dispersion curves for several lowest conduction-band states for a $D_3$ NW of the diameter $\approx 40$~\AA. The $\Gamma_4$ ($\Gamma_5 \oplus \Gamma_6$) states are shown by the green (black) lines. In agreement with Eq.~\eqref{eq:spin_split}, the states exhibit linear-in-$k$ spin splittings. The absolute values of $\alpha_{n}$, extracted from the dispersion calculated in tight-binding technique,
are shown in Fig.~\ref{fig:spin_splitting_r} as functions of the NW diameters for four lowest conduction band levels.
 Similar to the inter-valley splittings, the constants of the spin splitting strongly oscillate with the NW radius.

Due to large atomic spin-orbit interaction constants \cite{Poddubny12}
\[
	\Delta_{\text{Pb}} = 2.38~\text{eV},~~~\Delta_{\text{Se}} = 0.42~\text{eV},
\]
and a narrow band gap, the spin splitting constants may be huge (up to 1 eV$\cdot$\AA), 
as compared with typical values  
for A\textsubscript{3}B\textsubscript{5} 
quantum wells \cite{Jusserand92,Nestoklon16_110}. 
Although a special design of A\textsubscript{3}B\textsubscript{5}-based 
nanowires \cite{Soluyanov16} might lead to the same order of magnitude 
for the spin splitting, we stress that in our case we consider ideal 
nanowire of cylindrical shape based on material without bulk inversion asymmetry.

To make the spin splitting analysis complete, 
in Fig.~\ref{fig:spin_splitting_r} we show the spin splitting constants $\alpha_S$ as 
functions of the nanowire diameter. 
Each panel corresponds to a particular pair of states.
The panels are sorted by the longitudinal valley contributions, 
encoded by the color of points in Fig.~\ref{fig:spin_splitting_r}. 
One may note that the spin splitting is larger for the states with higher 
fractions of the inclined valleys.

\section{Conclusion}
\label{sec:conclusion}
In this work, we have calculated the energy spectrum of the 
[111]-grown PbSe nanowires in the framework of the empirical tight-binding method.
By comparing the atomistic results with the \KP\ theory we demonstrated
that the boundary scattering leads to the valley-orbit splitting of the 
confined states.
The valley splitting of the energy levels is very sensitive to a 
particular arrangement of atoms in the nanowire. 
We have considered PbSe nanowires grown along [111] direction
with $D_{3d}$, $D_{3}$ and $C_{2h}$ point group symmetry. 
Analysis of the local density of states in the reciprocal space allowed us to resolve the 
valley composition of the tight-binding wave functions. 
The strong correlation between the valley composition 
and the valley-orbit splittings shows that the main origin of the valley splitting is 
the intervalley mixing at the surface of the NW.

For relatively large nanowires with the diameters exceeding 40~\AA\ the energy
spectrum can be fairly well approximated in the framework of the \KP\ method,
but for the diameters less than $40$~\AA\ the valley splittings of confined
states become comparable with the energy distance between the unperturbed states 
from independent valleys and \KP\ theory should be modified to 
account for combinations of valley states.

The [111]-grown nanowires of the $D_{3}$ symmetry group represent a special case, 
as they lack inversion center and have a screw axis. For this reason they 
exhibit linear in wave vector spin-dependent splittings of energy levels. 
Our tight-binding calculations reveal giant  spin splitting constants
$\alpha$  up to $1$~eV$\cdot$\AA. 
This reflects the relativistic nature of electron spectrum in lead atoms 
and shows that the lead-chalcogenide nanowires are unexpectedly 
promising  candidates for the spintronic devices.

We believe that the intricate nature of the valley and spin splittings in small 
nanowires 
opens new opportunities in the control of spin and valley 
degrees of freedom.

\section*{Acknowledgments}
The authors acknowledge fruitful discussions with E.L.~Ivchenko. 
The work of IDA, ANP and MON was supported by the 
Russian Science Foundation (Grant No. 14-12-01067). 
The work of SVG was supported by
the National Science Foundation (NSF-CREST Grant HRD-1547754) 
and the Russian Foundation for Basic Research (Grant 15-02-09034).
 
\bibliography{PbSe}

\end{document}